# FALSE-ALARM PROBABILITY IN RELATION TO OVER-SAMPLED POWER SPECTRA, WITH APPLICATION TO SUPER-KAMIOKANDE SOLAR NEUTRINO DATA

PETER A. STURROCK[1] AND JEFFREY. D. SCARGLE[2]

[1]Center for Space Science and Astrophysics, Stanford University,
Stanford, CA 94305,USA
[2]NASA/Ames Research Center, MS 245-3, Moffett Field, CA 94035, USA

ABSTRACT

The term "false-alarm probability" denotes the probability that at least one out of M independent power values in a prescribed search band of a power spectrum computed from a white-noise time series is expected to be as large as or larger than a given value. The usual formula is based on the assumption that powers are distributed exponentially, as one expects for power measurements of normally distributed random noise. However, in practice one typically examines *peaks* in an oversampled power spectrum. It is therefore more appropriate to compare the strength of a particular peak with the distribution of peaks in oversampled power spectra derived from normally distributed random noise. We show that this leads to a formula for the false-alarm probability that is more conservative than the familiar formula. We also show how to combine these results with a Bayesian method for estimating the probability of the null hypothesis (that there is no oscillation in the time series), and we discuss as an example the application of these procedures to Super-Kamiokande solar neutrino data.

## 1. INTRODUCTION

For a power spectrum derived from normally distributed random noise, the probability that any one measurement will have the value S or more is given by the exponential distribution

$$P(\geq S) = e^{-S}. \qquad (1)$$

This quantity is known in statistics as a "p-value", that we write as

$$PV(S) = e^{-S}. \qquad (2)$$

Power-spectrum analysis typically takes no account of the phase of each frequency component, and we ignore phase considerations in this article.

The probability of finding the value S or more among M independent measurements, also a p-value, is referred to as the "false alarm probability" (Scargle, 1982), which we write as $FAP(S|M)$. Since the probability that all values are less than S is the product of the individual probabilities that each is less than S, we see that

$$FAP(S|M) \equiv PV(S|M) = 1 - \left(1 - e^{-S}\right)^M. \qquad (3)$$

If measurements are made at evenly spaced points, the number of independent measurements in a search band may be taken to be the bandwidth times the duration of the time series (which is dimensionless if both quantities are measured in the same units). If measurements are made at irregular intervals, it may be possible to estimate M from examination of properties of the time series, together with a prescription of the search band (Horne and Baliunas, 1986). In practice, analysts tend to over-sample a power spectrum to obtain an accurate estimate of the frequency of an oscillation in the time series. If there are M peaks in a prescribed search band and the biggest peak has power $S_P$, it is tempting to regard the peaks as independent measurements, and to use Equation (3) for the false-alarm probability, replacing S by $S_P$. (A reasonable check of the assumption that the peaks are independent is to subtract from the time series the sinusoidal function corresponding to the biggest peak, and then verify that the remaining peaks are substantially unaffected.)

However, there is a fallacy in this argument. We should not be comparing the value $S_P$ of a peak with the distribution of all possible power measurements derived from random noise. We should be comparing $S_P$ with the distribution of *peaks* to be expected in such a power spectrum. *When a power spectrum is over-sampled, the distribution of peaks is not the same as the distribution of powers in that power spectrum.* The mean value of the peak powers is—not surprisingly—substantially higher than the mean value of all of the powers. Hence the use of Equation (3) in this situation is misleading.

## 2. PEAK-POWER DISTRIBUTION

In order to apply the false-alarm concept to the peak powers in an over-sampled power spectrum, we need to determine the distribution of peak powers for power spectra derived from normally distributed random noise. If we can determine the distribution $D(S_P)$ of the peak powers, $S_P$, such that $D(S_P)\, dS_P$ gives the fraction of simulations for which $S_P$ is in the range $S_P$ to $S_P + dS_P$, we may arrive at an estimate of the p-value corresponding to a given value of a peak power from the expression

$$PV(S_P) = \frac{\int_{S_P}^{\infty} dx\, D(x)}{\int_{0}^{\infty} dx\, D(x)} . \qquad (4)$$

By analogy with the derivation of Equation (3), we see that the p-value associated with a peak of power $S_P$ among M independent peaks is given by

$$PV(S_P \mid M) = 1 - \left(1 - PV(S_P)\right)^{M} . \qquad (5)$$

If we introduce the quantity $S_P^*$ defined by

$$S_P^* = -\log(PV(S_P)), \qquad (6)$$

we see that the false-alarm probability for finding a peak of power $S_P$ among $M$ peaks is given by

$$FAP(S_P | M) \equiv PV(S_P | M) = 1 - \left(1 - e^{-S_P^*}\right)^M. \qquad (7)$$

Then the false-alarm probability associated with a peak of power $S_P$, among $M$ peaks, is the same as the false-alarm probability associated with a *value $S_P^*$ among M independent values* (not necessarily peaks).

## 3. MONTE CARLO ANALYSIS

It may be possible to determine the distribution of $D(S_P)$ analytically, but we have chosen to investigate it by a Monte Carlo procedure. We have generated a number of time series with times that are independently and randomly distributed with uniform probability, and a normal random distribution of measurements, and then formed the corresponding power spectra, using the Lomb-Scargle procedure (Lomb, 1976; Scargle, 1982). For each simulation, we have determined the values of the peaks in the power spectrum, defined as the frequencies at which the power is greater than both the powers at the two adjacent frequencies. From these simulations, we have determined the distribution $D(S_P)$ of the peak powers, $S_P$. We then determine the p-value for all values of SP using Equation (4), from which we may obtain $S_P^*$ using Equation (6). Figure 1 shows the resulting relationship between $S_P^*$ and $S_P$.

As we see from Figure 1, the following function is a good approximation to the empirically derived function:

$$S_P^* = \frac{S_P^{2.1}}{1.4 + S_P^{1.1}}.\qquad(8)$$

Apart from small values of $S_P$, the following is a fair approximation:

$$S_P^* \approx S_P - 1, \text{ for } S_P \geq 3.\qquad(9)$$

## 4. DISCUSSION

We now consider the question of how to fold the preceding results into our recent Bayesian approach to significance estimation (Sturrock and Scargle, 2009). Equation (7) gives the probability of obtaining a peak with value $S_P$ or more among $M$ independent peaks. This value is identical to the probability of obtaining the "equivalent" value $S_{EQ}$ at a single specified peak, if $S_{EQ}$ is chosen so that

$$e^{-S_{EQ}} = 1 - \left(1 - e^{-S_P^*}\right)^M,\qquad(10)$$

i.e. if

$$S_{EQ} = -\log\left(1 - \left[1 - e^{-S_P^*}\right]^M\right).\qquad(11)$$

We may now use the procedure in our recent article (Sturrock and Scargle, 2009) to compute $P(H0|S_{EQ})$. Equation (25) of that article yields the following expression for the odds on the null hypothesis (that there is no oscillation in the time series):

$$\Omega(H_0 | S_{EQ}) = 2.44\,(1.92 + S_{EQ})e^{-S_{EQ}}.\qquad(12)$$

The probability that $H_0$ is true is then given by

$$P(H_0) = \frac{\Omega(H_0)}{1 + \Omega(H_0)}.\qquad(13)$$

As an example, we consider the power spectrum (Sturrock et al. 2005a) derived from Super-Kamiokande data (Fukuda et al. 2001, 2002; Fukuda et al. 2003). The most prominent peak is found at 9.43 yr$^{-1}$ with power $S = 11.67$. For a search band 0 - 36 yr$^{-1}$ (the widest band consistent with no duplication due to aliasing), a count of the number of peaks gives $M= 126$. If we use the familiar form of the false-alarm probability given by Equation (3), we obtain $FAP(S|M) = 0.0011$. However, we find from Equation (9) that $S_P^* = 10.67$. Then Equation (7) leads to the estimate $FAP(S^*|M) = 0.0029$, which is somewhat more conservative than the earlier estimate.

Noting from Equations (11) that the false-alarm probability is the same as the p-value of a single peak with power $S_{EQ} = 5.84$, we find from Equations (12) and (13) that the probability that there is no oscillation in the time series is 0.052. This is considerably more conservative than either of the preceding estimates of the false-alarm probability, which is consistent with our recent conclusion (Sturrock & Scargle 2009) that one should never interpret a p-value as the probability that the null hypothesis is correct.

[We wish to point out that this discussion of the relationship of the concept of "false-alarm probability" to the Bayesian assessment of significance in power spectra supersedes that given in our recent article (Sturrock & Scargle, 2009). Our proposal (Equation (26) in that article) appeared to be analogous to the usual form of the false-alarm probability (Equation (3) of this article). However, Equation (26) of our preceding article does not lead to the same results as those based on Equations (11), (12) and (13) of the present article. Hence the apparent analogy has proved to be misleading.]

The false-alarm-probability concept can be applied to statistics which combine results from two or more power spectra, so designed as to have the same exponential distribution as a single power spectrum (Sturrock et al. 2005b). As an example, we consider the recent analysis of evidence for r-modes oscillations in Super-Kamiokande solar neutrino data (Sturrock 2008). We have examined r-mode frequencies as measured at Earth, given by

$$\nu(l,m,E) = m(\nu_R - 1) - \frac{2m\nu_R}{l(l+1)}, \qquad (14)$$

where $l$ ($\geq 2$) and $m$ $(1,2,...,l)$ are two of the three spherical-harmonic indices, and $v_R$ is the sidereal rotation frequency (in yr$^{-1}$). We have formed the joint power statistic (JPS; Sturrock et al., 2005b) from the powers measured at five frequencies corresponding to $m = 1$ and $l = 2, 3, 4, 5$ and $6$. For five power spectra, the joint power statistic is given, to sufficient accuracy, by

$$J = \frac{4.9 Y^2}{1.6 + Y} \quad (15)$$

where

$$Y = (S_1 \times S_2 \times ... \times S_5)^{1/5} . \quad (16)$$

We find a notable peak ($J = 11.48$) in the JPS spectrum at 13.97 year$^{-1}$. Equation (9) now leads to $J^* = 10.48$. In order to compare this result with our Monte-Carlo calculation (Sturrock 2008), we adopt 12 - 14 yr$^{-1}$ as the search band. We find that $J$ has 13 peaks in this band. On using $J^* = 10.48$ and $M = 13$ in Equation (10), we find that the p-value is 0.00037, not very different from the value 0.00027 that we derived from Monte Carlo simulations. Equation (10) also yields $S_{EQ} = 7.92$ as the equivalent power. On substituting this value in Equations (12) and (13), we arrive at the value 0.0086 for the probability of the null hypothesis (that the power spectrum is distributed exponentially, as in Equation (1)). This probability of the null hypothesis as derived from a Bayesian analysis is—as expected—significantly more conservative than the p-value (0.00027).

## Acknowledgements

One of us (PAS) acknowledges support from the NSF through Grant AST-0607572.

# REFERENCES

Fukuda, S., et al. 2001, Phys. Rev. Lett. 86, 5651

Fukuda, S., et al. 2002, Phys. Lett. B 539, 179

Fukuda, Y. 2003, Nucl. Instrum. Meth. Phys. Res. A 503, 114

Horne, J.H., & Baliunas, S.L. 1986, ApJ, 302, 757

Lomb, N.R. 1976. Astrophys. Space Sci., 39, 447

Scargle, J.D. 1982, ApJ 263, 835

Sturrock, P.A. 2008, Sol. Phys. 252, 221

Sturrock, P.A., & Scargle, J.D. 2009, ApJ, 706, 393

Sturrock, P.A., Caldwell, D.O., Scargle, J.D., & Wheatland, M.S. 2005a. Phys. Rev. D. 72, 113004

Sturrock, P.A., Scargle, J.D., Walther, G., & Wheatland, M.S. 2005b, Solar Phys. 227, 137

Willson, R.C. 1979, J. Applied Optics, 18, 179

Willson, R.C. 2001, The Earth Observer, 13, No. 3, 14

FIGURE

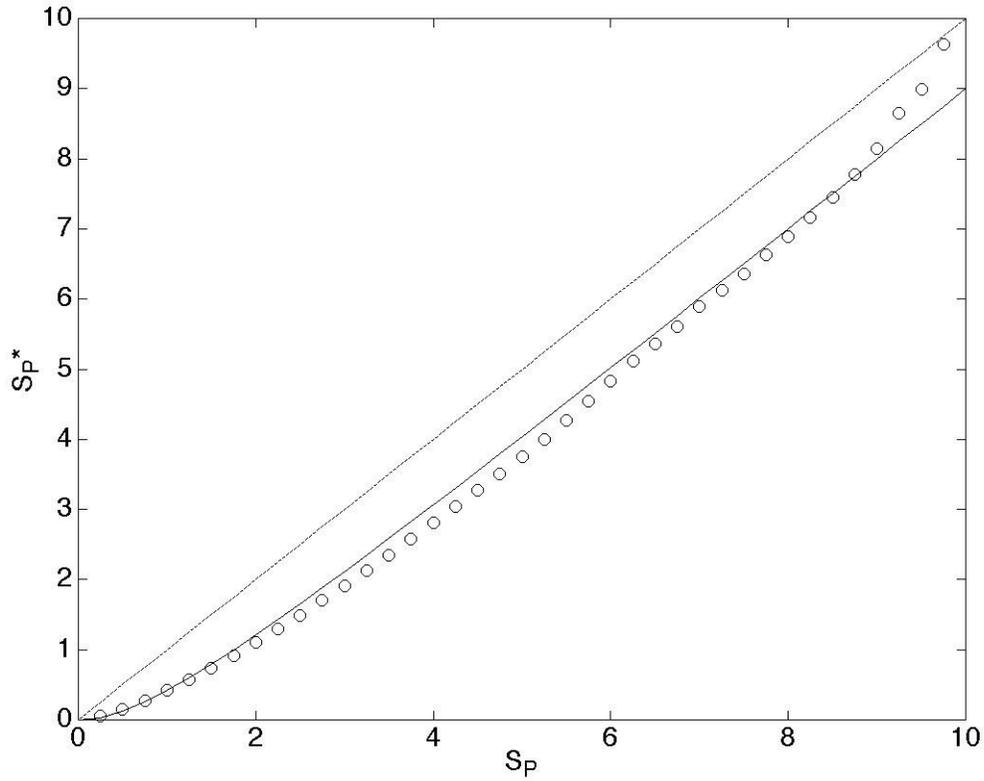

Figure 1. $S_P^*$ (shown as circles) as a function of $S_P$, derived from ~65,000 peaks in power spectra formed from 2,000,000 frequency samples of time series for which the times are random and measurements conform to a normal distribution. (The values of $S_P^*$ are not well determined for $S_P > 8$.) The fit from Equation (8) is shown as a solid line. The broken line shows the values of $S_P$ for comparison.